\documentstyle[11pt,newpasp,twoside,epsf]{article}
\begin{document}

\title{Kinematic Asymmetries in a Broad Galaxy Sample}
 
\author{S. J. Kannappan and D. G. Fabricant}
\affil{Harvard-Smithsonian CfA, 60 Garden St MS-20, Cambridge, MA 02138}

\begin{abstract}
We analyze asymmetries in the gas rotation curves of 113 galaxies
drawn from the Nearby Field Galaxy Survey, spanning a broad range of
luminosities and morphologies.  If the origin of the rotation curve is
chosen to minimize the asymmetry, then $\sim$23\% of sample galaxies
show asymmetry $>$5\% within 1.3$r_{e}$.  However, not all gas kinematic
asymmetries indicate asymmetries in the underlying potential.
\end{abstract}

\section{Introduction}
Roughly 50\% of bright spiral galaxies show asymmetry in their global
HI profiles (Haynes et al.\ 1998).  Such asymmetries may arise from
noncircular motions, lopsided gas distributions, or unresolved
companions.  By examining resolved optical rotation curves (RC's), we
can isolate the kinematic contribution to the asymmetry.  Although
optical RC's do not extend as far as {\em resolved} HI data, Courteau
(1997) points out that global HI profiles sample primarily the
kinematics of the inner disk, and high quality optical data generally
perform equally well.  We analyze only RC's extending to at least
1.3$r_{e}$, the peak velocity position for a pure exponential disk.
We also require that typical outer velocities reach $v_{\rm typ} > 30$
$\rm km \, s^{-1}$, deliberately introducing a bias against the
smallest and most face-on galaxies.  Although this bias is not in the
spirit of the minimum-bias parent survey (the Nearby Field Galaxy
Survey, Jansen et al.\ 2000), our velocity cut serves to focus
attention on galaxies with clear bulk motion, while still including a
wide range of types, and also luminosities as faint as M$_{\rm B} \sim
-15$.  The final sample comprises 113 galaxies, excluding one AGN.

\section{Asymmetry Measurements}
We adopt a quantitative asymmetry measure akin to the photometric
asymmetry index of Abraham et al.\ (1996).  Reflecting the rotation
curve about its origin, we compute the asymmetry within 1.3$r_{e}$ as
the average absolute deviation between the two sides, $<\mid\!v-v^{\rm
reflected}\!\mid>$.  (Asymmetries for $r>1.3$$r_{e}$ are more
difficult to compare due to the variable spatial extent of the data.)
Measured asymmetries depend critically upon the choice of origin, so
we shift the origin to numerically minimize the asymmetry.  We
constrain the spatial coordinate of the origin to remain within the
one-sigma error bars of our determination of the continuum peak
position, but we allow the velocity coordinate to vary freely.  The
final asymmetry is expressed as a percentage of the typical outer
velocity $v_{\rm typ}$.

Figure~1 shows the distribution of inner asymmetry values by
morphology.  About 23\% of our sample galaxies have asymmetry $>$5\%.
This number appears to be smaller than the HI result, consistent with
the idea that we are isolating kinematic asymmetries from lopsidedness
in the HI distribution, but direct comparison is difficult given the
differences in asymmetry measurement technique.  Some of the
asymmetries we see reflect peculiar gas dynamics with no stellar
counterpart, as illustrated in Figure~2, suggesting that the fraction
of galaxies with asymmetric gravitational potentials is probably
smaller than 23\%.

\begin{figure}
\plotone{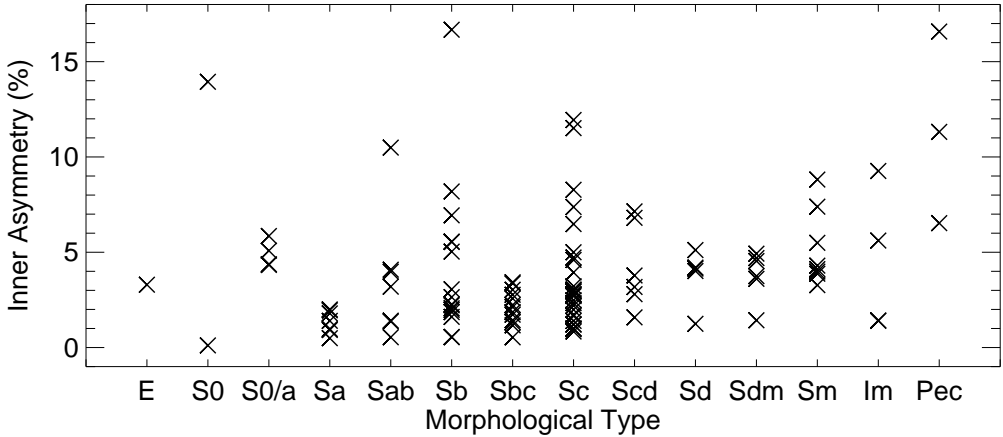}
\caption{Distribution of Asymmetry Values by Morphology.}
\end{figure}

\begin{figure}
\plotone{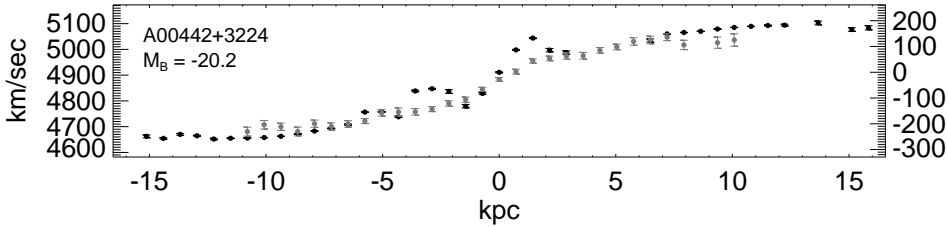}
\caption{Gas (black) and stellar (grey) RC's for a barred Sb.}
\end{figure}

\acknowledgements We thank R.~Jansen for sharing the NFGS photometry.
S.~J.~K. received support from a NASA GSRP Fellowship.

\end{document}